\documentclass{article}
\usepackage{spconf,amsmath,epsfig}

\let\OLDthebibliography\thebibliography
\renewcommand\thebibliography[1]{
  \OLDthebibliography{#1}
  \setlength{\parskip}{0pt}
  \setlength{\itemsep}{0pt plus 0.3ex}
}

\usepackage{amsfonts}  
\usepackage{amssymb}
\usepackage{bm}

\begin{document}\sloppy

\def\x{{\mathbf x}}
\def\L{{\cal L}}

\title{Hierarchical Information Enhancement Network for Cascade Prediction in Social Networks}
%
\name{Fanrui Zhang\textsuperscript{\rm 1}, Jiawei Liu\textsuperscript{\rm 1}, Qiang Zhang\textsuperscript{\rm 1},  Xiaoling Zhu\textsuperscript{\rm 2}, Zheng-Jun Zha\textsuperscript{\rm 1}}
\address{\textsuperscript{\rm 1} University of Science and Technology of China, China\\
                    \textsuperscript{\rm 2} CETC Academy of Electronics and Information Technology Group Co, China\\
                    \{zfr888, zq\_126\}@mail.ustc.edu.cn\\
                    zhuxiaoling1@cetc.com.cn\\
                    \{jwliu6, zhazj\}@ustc.edu.cn
                    }

\maketitle

\begin{abstract}
Understanding information cascades in networks is a fundamental issue in numerous applications. Current researches often sample cascade information into several independent paths or subgraphs to learn a simple cascade representation. However, these approaches fail to exploit the hierarchical semantic associations between different modalities, limiting their predictive performance. In this work, we propose a novel Hierarchical Information Enhancement Network (HIENet) for cascade prediction. Our approach integrates fundamental cascade sequence, user social graphs, and sub-cascade graph into a unified framework. Specifically, HIENet utilizes DeepWalk to sample cascades information into a series of sequences. It then gathers path information between users to extract the social relationships of propagators. Additionally, we employ a time-stamped graph convolutional network to aggregate sub-cascade graph information effectively. Ultimately, we introduce a Multi-modal Cascade Transformer to powerfully fuse these clues, providing a comprehensive understanding of cascading process. Extensive experiments have demonstrated the effectiveness of the proposed method.
\end{abstract}
\section{Introduction}
\label{sec:intro}

The advent of social media has fundamentally altered the traditional ways in which people acquire information online. Users enjoy the convenience and efficiency of sharing information and exchanging opinions on online social platforms \cite{jin2022adaptive,tatar2014survey}. As individuals share content with their contacts, these messages can be further amplified through their networks, sparking a chain reaction known as information cascades \cite{wu2022price,bikhchandani1992theory}. A deep understanding of these cascades is vital due to their significant influence on economic, political, and societal landscapes \cite{wu2023pda,dong2015will}. The prediction of a message's reach, or its popularity, has thus become a focal point of interest across both academic research and commercial endeavors \cite{ji2023relationship,pinto2013using}. 

In recent years, a number of studies have concentrated on this area, with a particular emphasis on the popularity prediction of items within social networks. Current approaches to cascade popularity prediction fall into three main categories: 
(1) Feature-based methods: Scholars have directed their efforts toward the extraction and amalgamation of bespoke features to forecast information cascades. These methodologies \cite{wu2023pda,yang2022bmp,cao2017deephawkes} hinge on an array of features such as the content's characteristics, user profiles, network structures, and temporal dynamics. Despite their reliance on in-depth domain expertise, these models may falter in predicting cascades within contexts that are new or not well-understood. (2) Generative methods: These methods describe the social phenomena in real networks as a series of continuous event sequences \cite{shen2014modeling,perozzi2014deepwalk}. They view the accumulation of information propagation as a process of forwarding arrivals and focus on independently modeling the forwarding rate function of each message \cite{gao2016modeling,mishra2016feature,zhao2015seismic}. However, generative methods lack supervision for future information prediction, leading to suboptimal performance in predicting cascade information diffusion. (3) Deep learning-based methods: Recent advancements in deep learning have achieved significant success in various applications. Researchers leverage various deep learning techniques and develop models to capture the temporal and sequential processes of information diffusion. Most methods represent cascade graphs as several paths through random walks \cite{perozzi2014deepwalk} or by constructing subgraphs of cascade diffusion \cite{cascn,zhu2023casciff}. Despite significant advances in information cascade prediction through deep learning methods, there remain inherent limitations. Primarily, these methods often model cascades from a singular perspective, either representing cascade graphs as paths learned through random walks \cite{li2017deepcas} or constructing graphs with relationships, typically overlooking the complementary information between these modalities. Additionally, different social platforms possess unique user behaviors and friend structures; learning to encode these social graph relationships can be crucial for uncovering the latent influence of propagators. Lastly, existing multi-modal feature fusion in cascade prediction typically relies on simple concatenation or attention mechanisms \cite{vaswani2017attention}, which do not effectively integrate these diverse multi-modal cues.

In this work, we propose a novel Hierarchical Information Enhancement Network (HIENet) for cascade popularity prediction in social networks, aimed at predicting the number of users influenced by a piece of information. It jointly models three types of modal features: social graph information, sub-cascade graph information, and cascade sequence information, learning an effective hierarchical feature-enhanced cascade representation. Specifically, HIENet comprises three multi-modal information processing modules, a Multi-modal Cascade Transformer, and a classifier.
1) In multi-modal information processing modules, we incorporate the extraction process for three types of modal information. 
In the cascade sequence information processing, we first sample the cascade graph into several propagation path sequences of length $N$ to mine and learn basic information propagation features. Subsequently, each path sequence is represented as a node, and a bidirectional LSTM \cite{schuster1997bidirectional} is used to encode interactions among these paths. 
In the sub-cascade graph information processing, we organize the propagation graph into a sequence of sub-cascades rather than propagation paths. Each sequence contains a series of sub-graphs, with the first sub-graph containing only the information's starting point and each subsequent sub-graph adding one more forward. To incorporate time information, we use the forwarding time's positional encoding as the sub-graph's information and employ a Graph Convolutional Network (GCN) \cite{kipf2016gcn} to extract the topological structure of the propagation graph. 
In Social graph information processing, we encode the latent influence of each information propagator by designing a users relevance reasoning strategy to find the shortest semantically relevant path between each pair of users in the global cascade graph and absorb all complementary information of the connected users in this path to learn feature-enhanced social graph information.
2) After obtaining feature information from the three modalities, our Multi-modal Cascade Transformer introduces a set of learnable token [CAS] along with the three feature inputs into the transformer for feature fusion, where the learnable set of token [CAS] makes the heterogeneous features consistent in a common feature space. 
3) Ultimately, the learnable token labeled [CAS], which encapsulates the enhanced modal information, is input into a classifier (composed of a Multi-Layer Perceptron) tasked with the prediction of cascade popularity.
Overall, our main contributions can be summarized as follows:
(1) We explore and obtain three beneficial features for cascade prediction from multiple dimensions of the cascade information.
(2) In Social graph information processing, we encode each information propagator's latent influence with a user-enhanced paths measurement strategy.
(3) The proposed Multi-modal Cascade Transformer effectively fuses multi-modal cascade information, reducing modality disparity.
(4) Comparative experiments on two real social platform datasets show that the method has good prediction results.

\begin{figure*}[t]
    \centering
    \includegraphics[width=1\textwidth]{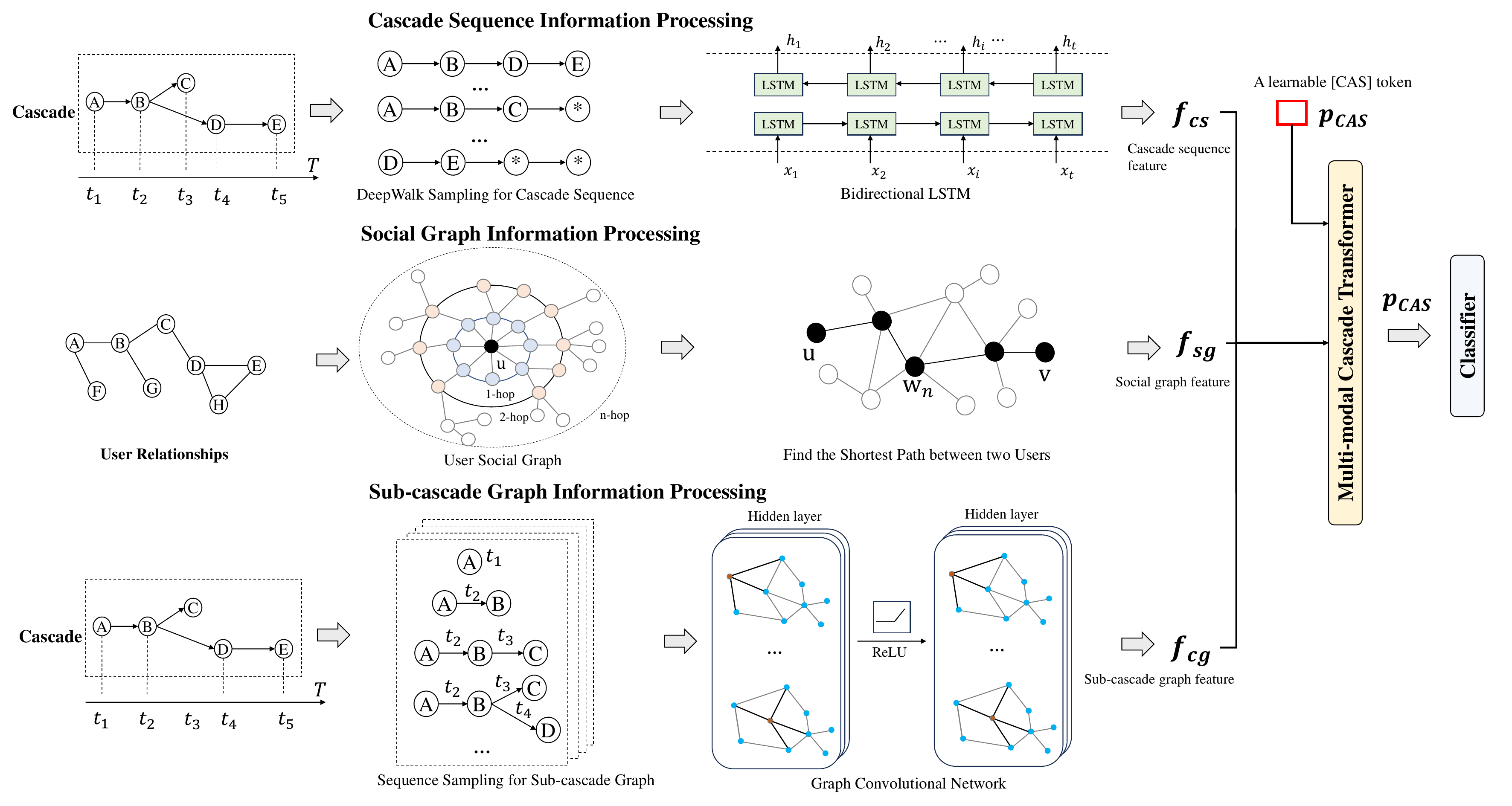}
    \caption{The overall architecture of the proposed HIENet. It mainly contains five modules: Cascade sequence information processing, Social graph information processing, Sub-cascade graph information processing, Multi-modal Cascade Transformer and Classifier. The left side of the figure shows the raw cascade information.}
    \label{fig:overoframework}
\end{figure*}

\begin{figure*}[t]
    \centering
    \includegraphics[width=1\textwidth]{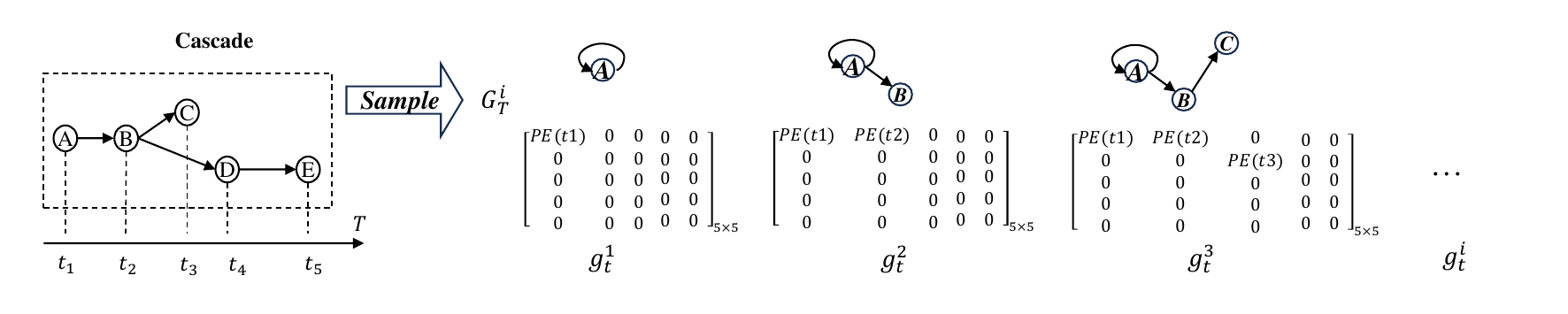}
    \caption{Illustration of sampling and representation of sub-cascade graph sequence.}
    \label{fig:sample}
\end{figure*}

\section{Method}
As shown in Figure ~\ref{fig:overoframework}, our Hierarchical Information Enhancement Network mainly consists of several parts: cascade sequence information processing, Social graph information processing, Sub-cascade graph information processing, Multi-modal Cascade Transformer and Classifier.
The details about the parts are described in the following subsections. 

\subsection{Problem Definition}
Existing work typically regards the task of cascade popularity prediction as a regression problem, that is, predicting the precise future popularity.
A cascade graph $G=(U,E,T)$ represents the diffusion process of messages over time with users $U$, relationships $E$, and timestamps $T$. For each message $m^i$, a cascade $C^i$ is constructed to track the diffusion process, comprising tuples that denote the retweeting user, the original poster, and the time elapsed between the retweets.
For example, in Figure ~\ref{fig:overoframework}, the cascade can be represented as \{(A, $t_1$ = 0), (A, B, $t_2$ ), (B,C, $t_3$),
(B, D, $t_4$), (D, E, $t_5$ )\}. Given the cascades in the observation
time window [0,T), we aim to predict the incremental popularity $S_{\Delta}^i$
between observed popularity $S_{T}^i$ and final popularity of each cascade $C^i$.

\subsection{Cascade Sequence Information Processing}
In our approach, we adopt a probabilistic method for selecting the starting node for a walk in the cascade graph, which is reminiscent of the methodology employed by DeepWalk \cite{perozzi2014deepwalk,li2017deepcas}. We define the probability of choosing a particular node 
$v$ as the starting node as follows:
\begin{align}
P(v) = \frac{\text{deg}_{+}(v) + \beta}{\sum_{w \in V} (\text{deg}_{+}(w) + \beta)}
\end{align}
where $\text{deg}_{+}(v)$ is the out-degree of node $v$, $V$ is the set of all nodes in the graph, and $\beta$ is a smoothing parameter to prevent the domination of high-degree nodes.
The transition probabilities to the next nodes are as follows:
\begin{align}
p(u | v) = \frac{\text{deg}_{+}(u) + \beta}{\sum_{w \in N(v)} (\text{deg}_{+}(w) + \beta)}
\end{align}
where $N(v)$ denotes the set of out-degree neighbors of 
$v$, ensuring that each path extends to length 
$N$ or receives padding with a designated symbol when no further nodes are available. 
Building upon this foundation, we employ a decision mechanism for random jumps and transitions to neighbor nodes, which facilitates the automatic determination of the sequence count 
$k_v$ and length $t_c$ that best capture the characteristics of each specific cascade graph. Consequently, we obtain 
$K$ sequences of length $N$. These sequences are then processed as hidden states through a bidirectional LSTM \cite{schuster1997bidirectional}, which encodes the temporal and structural dynamics of cascades, yielding a comprehensive feature representation $f_{cs}$ of the cascade sequences. 
For each node sequence $ X = (x_1, x_2, \ldots, x_N) $, the LSTM updates are follows:
\begin{align}
f_{cs} = \text{BiLSTM}(h_{t-1}, x_t)
\end{align}
where $x_t$ is the input feature at time step $t$, and $h_t$ is the hidden state combining both the forward and backward information at time $t$. The final feature representation $f_{cs}$ is then obtained by aggregating these hidden states.

\subsection{Social Graph Information Processing}

Given users in a cascade graph, we propose a novel metric $\bm{D}(u, v)$ to measure the augmented distance between two users on the global social graph. Unlike previous approaches that directly calculate feature distances between users without considering propagation paths, the $\bm{D}$ metric is able to utilize feature distances in the embedding space as well as graph distances on the global social graph topology, which is more suitable for path correlation modeling. We first find a shortest correlation path $\pi$ connecting $u$ and $v$ in the social graph:
\begin{align}
  \pi :  u = {w_0} \rightarrow {w_1} \rightarrow \ldots \rightarrow {w_n} ={v}
\end{align}
Where $n$ denotes the number of the entities in $\pi$. After obtaining the optimal path $\pi$, the module refines the representation ${E}^u$ for user $u$: 
\begin{align}
  {E}^u=\frac{1-\alpha}{1-\alpha^{n+1}}\sum_{i=0}^n \alpha^i {g}^{w_i}
\end{align}
where ${g}^{w_i}$ is the feature embedding of user ${w_i}$ in the path. $\alpha$ denotes a weight coefficient. It's worth noting that $ {E}^u$ is a path-aware representation, \textit{i.e.}, a different pair $(u, {v'})$ with path $\pi'$ will yield a different value of ${E}^u$. 
We leverage the enriched user features from various paths to acquire the social graph feature.
\begin{align}
f_{sg} = \text{concat}({E}^{u_1}, {E}^{u_2}, \ldots, {E}^{u_n})
\end{align}

\subsection{Sub-cascade Graph Information Processing}

Different from the cascade sequence information processing, this module constructs a sequential array of sub-cascade graphs, expressed as $\{g_t^1, g_t^2, \ldots, g_t^m\}
$, to encapsulate the evolving structural dynamics from one stage of the cascade to the next \cite{cascn}. Each graph in this sequence not only emanates from its antecedent but also assimilates and amplifies its influence, depicting an ongoing, cumulative progression. Crucially, the initial trajectory of information spread is a vital factor for its eventual scale, making the extraction of insights from the nascent phase of diffusion a requisite for understanding the full scope of propagation.
Moreover, as shown in Figure ~\ref{fig:sample}, to capture the temporal dynamics of user influence in information diffusion, we employ positional encoding to delineate the specific moments of user activations. After embedding the timestamps into a unified vector space, at a given time $t$, the encoding is designed to uniquely identify each time step, assigning it a distinct temporal signature. The encoding for time $t$ is defined as follows:
\begin{align}
PE(t)_{2d} = \sin\left(\frac{t}{10000^{2d/D}}\right) \\
PE(t)_{2d+1} = \cos\left(\frac{t}{10000^{2d/D}}\right)
\end{align}
where $1 \leq d \leq D/2$ denotes the dimension in the temporal embedding $PE(t)$.
$D$ denotes the total span of timestamps embedding from the beginning to the end of the cascade. 

Finally, we employ a two layers graph convolutional network (GCN) to aggregate information from cascade snapshots, obtaining the final sub-cascade graph features denoted by $ f_{cg} $.
\begin{align}
f_{cg} = \sigma(\tilde{D}^{-\frac{1}{2}} \tilde{A} \tilde{D}^{-\frac{1}{2}} H W)
\end{align}
where $\tilde{A} = A + I$ is the adjacency matrix with self-connections, $I$ is the identity matrix. $\tilde{D}$ is the degree matrix of $\tilde{A}$; $W$ is the weight matrix; $H$ is the feature matrix.

\subsection{Multi-modal Cascade Transformer}

The Multi-modal Cascade Transformer module consists of three linear projection layers (one for each modality) and a transformer layer.
After obtaining the features for the three modalities, we apply the linear projection layers to project them into embeddings with the same dimensions. We denote the projected cascade sequence feature, social graph feature and sub-cascade graph feature embeddings as $\bm{f}_{cs}$,$\bm{f}_{sg}$ and $\bm{f}_{cg}$, respectively, and then, we pre-add a learnable embedding (i.e., [CAS] token) and formulate the joint input feature of the transformer layer as:
\begin{equation}
	\bm{x}_{input} = [f_{cs}, f_{sg},\ f_{cg}, p_{CAS}\ ],
\end{equation}
where $p_{CAS}$ represents the [CAS] token. The [CAS] token is randomly initialized at the beginning of the training stage and optimized with the whole model.
After obtaining the input $\bm{x}_{input}$ in the above joint embedding space, we apply the transformer layer to embed $\bm{x}_{input}$ into a common semantic space by performing intra- and inter-modal relational reasoning. The output states of [CAS] token form a comprehensive representation enriched by the three modalities and are further used for cascade prediction. 

\subsection{Classifier}
We leverage the output state of [CAS] token as the input of our Classifier. 
\begin{equation}
S_{\Delta}^i = \text{MLP}([CAS])
\end{equation}
where $MLP$ is a multi-layer perceptron.
The primary aim is to reduce the loss function as follows:
\begin{equation}
\mathcal{E}(S_{\Delta}^i, \hat{S}_{\Delta}^i) = \frac{1}{N} \sum_{i=1}^{N} (\log S_{\Delta}^i - \log \hat{S}_{\Delta}^i)^2,
\end{equation}
where $N$ denotes the total number of messages, 
$S_{\Delta}^i$ denotes the predicted incremental size for the specific message 
, and $\hat{S}_{\Delta}^i$ represents the observed incremental size.

\begin{table*}[t]
\caption{ Performance comparison to the state-of-the-art methods on Weibo and APS datasets.}
  \centering
  \footnotesize
  \resizebox{2.0\columnwidth}{!}{
  \renewcommand{\arraystretch}{1.0}
\begin{tabular}{l|cc|cc|cc|cc|cc|cc}
\hline
Dataset & \multicolumn{6}{c|}{Sina Weibo} & \multicolumn{6}{c}{APS} \\
\hline
T & \multicolumn{2}{c|}{1 hour} & \multicolumn{2}{c|}{2 hours} & \multicolumn{2}{c|}{3 hours} & \multicolumn{2}{c|}{5 years} & \multicolumn{2}{c|}{7 years} & \multicolumn{2}{c}{9 years} \\ 
\hline
Metric & MSLE & mSLE & MSLE & mSLE & MSLE & mSLE & MSLE & mSLE & MSLE & mSLE & MSLE & mSLE \\

\hline
SEISMIC & --- & 0.657 & --- & 1.084 & --- & --- & 1.575 & 0.874 & --- & 0.717 & --- & 0.712 \\
Feature-linear & 3.701 & 1.058 & 3.365 & 1.105 & 3.328 & --- & 1.582 & 0.679 & 1.508 & 0.674 & 1.456 & 0.722 \\
DeepCas & 3.631 & 0.808 & 3.213 & 0.837 & 3.107 & 0.902 & 1.629 & 0.671 & 1.538 & 0.603 & 1.462 & 0.662 \\
DeepHawkes & 2.448 & 0.650 & 2.279 & 0.675 & 2.223 & 0.639 & 1.510 & 0.652 & 1.337 & 0.584 & 1.211 & 0.605 \\
CasCN & 2.395 & 0.661 & 2.230 & 0.745 & 2.223 & 0.699 & 1.489 & 0.623 & 1.354 & 0.594 & 1.310 & 0.602 \\
CasFlow & 2.241 & 0.525 & 2.405 & \textbf{0.569} & 2.400 & 0.605 & 1.373 & 

\textbf{0.524} & 1.344 & 0.573 & 1.353 & 0.595 \\
I3T & 2.182 & 0.729 & 2.190 & 0.738 & 2.105 & 0.717 & 1.361 & 0.592 & 1.273 & 0.564 & 1.186 & 0.582 \\
HIENet & \textbf{2.178} & \textbf{0.521} & \textbf{2.169} & 0.615 & 

\textbf{2.031} & \textbf{0.601} & \textbf{1.291} & 0.571 & \textbf{1.204} & \textbf{0.554} & \textbf{1.121} & \textbf{0.545} \\
\hline

\end{tabular} 
}
\label{tab:comparation}
\end{table*}

\section{Experiments}
\subsection{Experimental Settings}
\textbf{Dataset.}
Experiments were conducted in two distinct application scenarios, Sina Weibo and the APS datasets, to assess the efficacy of the proposed HIENet model. Sina Weibo \cite{cao2017deephawkes} serves as a real-time social media platform for information sharing, with the dataset sourced from DeepHawkes\cite{cao2017deephawkes}. The number of retweets within 24 hours approximates the ultimate popularity, making the 24-hour mark post-publication the chosen prediction time. Observation times are set at 1, 2, and 3 hours post-release.
The APS dataset, a public repository, encompasses research papers published by the American Physical Society from 1893 to 2009, featuring 463,344 papers, 245,365 authors, across 19 journals, and accruing 4,692,026 citations. In the cascade graph, an edge points from paper A to B if A cites B. The observation windows are set at 5, 7, and 9 years post-publication, with the 20th year as the benchmark for predicting ultimate citation counts. Our objective is to forecast future increments in paper citations.

\textbf{Implementation Details.} For the random walks, each cascade graph generates 100 sequences with a maximum length of 20. The Bi-LSTM utilizes a hidden layer dimension of 256. The MLP comprises two fully connected layers with sizes of 128 and 32, respectively. The learning rate is set to 1e-4. $\beta$ is set to 0.8. $\alpha$ is set to 0.9. Considering the characteristic power-law distribution of message popularity on social networks, we opt for the mean square logarithmic error (MSLE) as a more fitting performance metric rather than the conventional mean square error. Furthermore, to assess the median level of model inaccuracy, we employ the median of the square log-transformed errors, denoted as mSLE.

\begin{table}[!t]
	\caption{Evaluation of the influence of different modules of HIENet on Weibo dataset.}
	\begin{center}
	\begin{tabular}{l|ccc}
\hline
Dataset & \multicolumn{3}{c}{Sina Weibo} \\
\hline
\( T \) (hours) & 1 & 2 & 3 \\
\hline
w/o cascade sequence feature & 3.261 & 3.534 & 3.117 \\
w/o social graph feature & 2.585 & 2.396 & 2.325 \\
w/o sub-cascade graph feature & 2.685 & 2.448 & 2.699 \\
w/o cascade transformer & 2.385 & 2.248 & 2.236 \\
HIENet & \textbf{2.178} & \textbf{2.169} & \textbf{2.031} \\
\hline
\end{tabular}
		\label{tab: Ablation Study_1}
	\end{center}
\end{table}

\subsection{Comparison to State-of-the-Art Approaches}
In order to evaluate the HIENet, we compare it with the following state-of-the-art methods on the Weibo and APS datasets, including SEISMIC \cite{zhao2015seismic}, Feature-linear \cite{cao2017deephawkes}, DeepCas \cite{li2017deepcas}, DeepHawkes \cite{cao2017deephawkes}, CasCN \cite{cascn}, CasFlow \cite{xu2021casflow} and $I^3T$ \cite{i3t}. In Table ~\ref{tab:comparation}, we can observe that the proposed HIENet achieves the best performance respectively on the two datasets. 
The results demonstrate the effectiveness of the proposed method. 
Among these compared methods, Feature-based methods (i.e., Feature-linear) perform poorly due to extensive manual feature extraction and a lack of effective universal features. 
The generative method SEISMIC underperforms in cascade prediction as it lacks future popularity guidance.
DeepHawkes and DeepCas achieved notable performance by sampling cascade graphs into multiple propagation path sequences to learn dissemination behaviors. CasCN organizes propagation graphs into sequences of sub-cascade graphs, encompassing a series of subgraphs. CasFlow excelled by further refining and learning node representations within independent sub-cascade graphs. $I^3T$ attained commendable performance by encoding inter- and intra-cascade path features, as well as emphasizing temporal characteristics.
Compared to the aforementioned methods, we can attribute the strength of HIENet to several aspects:
1) We extract diverse features from three modalities: sub-cascade graph, social graph, and cascade sequence, achieving comprehensive mining and understanding of cascade graphs.
2) In the processing of sub-cascade graph information, we enhance the model's comprehension and extraction of temporal features by introducing time positional encoding within various subgraph sequences.
3) We capture the social graph features of users at the inception of cascade propagation. By amalgamating the user and propagation data across diverse trajectories, we significantly enhance the robustness of the model.

\subsection{Ablation Studies}

\textbf{Analysis of detailed features.} The results in Table \ref{tab: Ablation Study_1} show the influence of different features. The three columns from top to bottom correspond to without the three features in Figure \ref{fig:overoframework}:
the cascade sequence feature $\bm{{f}_{cs}}$, the social graph feature $\bm{{f}_{sg}}$ and the sub-cascade graph feature $\bm{{f}_{cg}}$. 
We demonstrate the benefits of each feature, which highlights the importance of integrating all modalities for cascade prediction. Removing the cascade sequence feature $\bm{{f}_{cs}}$ or the sub-cascade graph feature $\bm{{f}_{cg}}$ significantly reduces performance, which demonstrates the importance of these two features. 

\textbf{Analysis of Multi-modal Cascade Transformer.} We conduct experiments to analyze the effectiveness of the proposed Multi-modal Cascade Transformer. \textit{w/o Cascade Transformer} denotes HIENet without using the Multi-modal Cascade Transformer for feature fusion (replacing with concatenate operation). 
We observe that the Multi-modal Cascade Transformer contributes to performance improvement. The comparative results highlight the advantages of projecting both features into the same vector space, which facilitates a comprehensive feature fusion.

\section{Conclusion}
In this work,  we propose a novel Hierarchical Information Enhancement Net-
work (HIENet) for cascade prediction in social networks, which jointly models three types of modal features: social graph information, sub-cascade graph information, and cascade sequence information.
Concretely, we design a novel Multi-modal Cascade Transformer to facilitate a comprehensive feature fusion.
Extensive experiments on two benchmarks \textit{i.e.}, Sina Weibo and APS datasets validate the effectiveness of HIENet.

\bibliographystyle{IEEEbib}
\bibliography{icme2023template}

\end{document}